\documentclass[aps,showpacs, nofootinbib]{revtex4}

\newcommand{\be}{\begin{equation}}
\newcommand{\ee}{\end{equation}}
\newcommand{\ben}{\begin{eqnarray}}
\newcommand{\een}{\end{eqnarray}}

\newcommand{\cL}{{\cal L}}
\newcommand{\cH}{{\cal H}}

\newcommand{\p}{\partial}
\newcommand{\na}{\nabla}

\newcommand{\ep}{\epsilon}
\newcommand{\bep}{\bar \epsilon}

\newcommand{\bdel}{\bar \delta}

\newcommand{\ga}{\gamma}


\begin{document}

\title{First Law of Black Rings Thermodynamics in Higher Dimensional Chern-Simons Gravity}

\author{Marek Rogatko}
\affiliation{Institute of Physics \protect \\
Maria Curie-Sklodowska University \protect \\
20-031 Lublin, pl.~Marii Curie-Sklodowskiej 1, Poland \protect \\
rogat@tytan.umcs.lublin.pl \protect \\
rogat@kft.umcs.lublin.pl}

\date{\today}

\begin{abstract}
The {\it physical process} version and the {\it equilibrium state} version of the first law
of black ring thermodynamics in $n$-dimensional Einstein gravity with Chern-Simons term
were derived. This theory constitutes the simplest generalization of the five-dimensional
one admitting a stationary black ring solutions. The {\it equilibrium state} version of the first law
of black ring mechanics was achieved by choosing any cross section
of the event horizon to the future of the bifurcation surface.

\end{abstract}

\maketitle

\section{Introduction}
Various attempts to unify the forces of Nature involve a great resurgence of the consideration
of spacetimes of dimensionality greater than four. Consequently, there occurs also continuously growing interest
in $n$-dimensional black holes and its mathematical structure. Namely, the uniqueness theorem for 
static $n$-dimensional black holes was quite well established \cite{uniq}. Recently,
the proof of the rigidity theorem in higher dimensional gravity for non-extremal black holes was
provided \cite{hol06}. On the other hand, for stationary axisymmetric $n$-dimensional solutions
the situation is far from obvious. It was shown \cite{emp02} that even in five-dimensional
spacetime there is the so-called {\it black ring} solution having $S^2 \times S^1$
topology of the event horizon. It has the same mass and angular momentum as a spherical 
five-dimensional stationary axisymmetric black hole. If one assumes the topology of
black hole event horizon as $S^3$ the uniqueness
proof can be established (see Ref.\cite{mor04} for the vacuum case and Ref.\cite{rog04a} for the 
stationary axisymmetric self-gravitating $\sigma$-model).
In literature, there has been black ring solution possessing both electric and magnetic dipole
charges \cite{elv03,elv03b}, also static black ring solution has been found in 
five-dimensional Einstein-Maxwell-dilaton gravity \cite{kun05} 
and systematically derived in \cite{yaz05} both in asymptotically
flat and non-asymptotically flat case. There are also a supersymmetric generalizations
of these objects \cite{sup} (for a review of a black ring story see \cite{emp06} and references therein).
\par
As the interest of these object systematically grows we shall in our paper
study the first law of black ring thermodynamics in the case of $n$-dimensional
gravity coupled to Chern-Simons (CS) term. In five-dimensional case
of Einstein gravity with CS term, i.e., a minimal supergravity theory, 
the black ring solution was given in Ref.\cite{elv05}. We shall consider the simplest
generalization of this theory to the case of $n$-dimensions. First, we look for
the {\it physical process} version of the first law of black ring thermodynamics.
In black hole physics it was realized changing a stationary black hole by some infinitesimal physical process
, e.g., by throwing matter into black hole. When one assumes that the 
final state of black hole settles down to a 
stationary one, we can extract the changes of black hole's parameters and 
in this way conceive the idea about the first law of black hole mechanics. 
The {\it physical process } version of the first 
law of black hole thermodynamics was studied in Einstein and Einstein-Maxwell (EM)
theory in Refs.\cite{wal94,gao01} and
in
Einstein-Maxwell axion-dilaton (EMAD) gravity being the low-energy limit of the heterotic
string theory in Ref.\cite{rog02}. Having in mind the assumption about spherical topology
of $n$-dimensional black hole the {\it physical process} version of the first law of thermodynamics
was treated in Ref.\cite{rog05} in the case of Einstein gravity coupled to
$(n-2)$-gauge form field strength. This kind of derivation of the first law of thermodynamics
was also elaborated in the case black rings in higher dimensional gravity containing
$(p+1)$-form field strength and dilaton field \cite{rog05br}, being
the simplest generalization of five-dimensional one in which stationary black ring
solution has been provided \cite{elv05}.
\par
The other attitude to the problem of the first law of black hole thermodynamics is the so-called
{\it equilibrium state} version. It was studied in the seminal paper of
Bardeen, Carter and Hawking \cite{bar73}. This attitude is based on taking into account
the linear perturbations of a stationary electrovac black hole to another one.
Arbitrary asymptotically flat perturbations of a stationary 
black hole were considered  in Ref.\cite{sud92}, while the first law of black hole thermodynamics valid for
an arbitrary diffeomorphism invariant Lagrangian 
with metric and matter fields possessing stationary and axisymmetric
black hole solutions were given in Refs.\cite{wal93}-\cite{iye97}. The cases of higher 
curvature terms and higher derivative terms in the metric
were considered in
\cite{jac}, while the situation when the Lagrangian is an arbitrary function of metric,
Ricci tensor and a scalar field was elaborated in Ref.\cite{kog98}.
In Ref.\cite{gao03}
of a charged rotating black hole
where fields were not smooth through the event horizon was treated.
\par
The first law of black hole thermodynamics was also provided in the case of $n$-dimensional
black holes. The {\it equilibrium state} version was studied in Ref.\cite{equi1} .
Some of the works assume that four-dimensional black 
hole uniqueness theorem extends to higher dimensional case are devoted to the problem \cite{equi2}.\\
In Ref.\cite{cop05}, the authors using the notion of
bifurcate Killing horizons and taking into account dipole charges were managed
to find the first law of black hole thermodynamics for black ring solutions.
In the higher dimensional gravity containing $(p + 1)$-form field strength and dilaton fields
the first law of black ring mechanics choosing an arbitrary cross
section of the event horizon to the future of the bifurcation surface was derived in Ref.\cite{rog05br1}.
\par
Our paper will be devoted to the first law of black ring thermodynamics in $n$-dimensional
Einstein CS gravity, being the simplest generalization of a five-dimensional theory which
admits stationary black ring solution. In Sec.II we shall elaborate 
the {\it physical process} version of the first law of black ring thermodynamics, while
Sec.III will be devoted to the {\it equilibrium state} version of the first law. We shall
derive this law by choosing the arbitrary cross section of the black ring event horizon
to the future of the bifurcation surface. It allows one to treat fields which are not necessary 
smooth through the event horizon. Our requires only that the pull-back of the fields 
in the future of bifurcation surface be smooth.

\section{Physical process version of the first law of black ring mechanics}
In this Sec. we shall consider the simplest higher dimensional generalization a
of minimal five-dimensional supergravity theory where a stationary black ring solutions have been found.
In five-dimensional case the solutions were sufficiently complicated that a first
law of thermodynamics could not be found by inspection \cite{elv05}. In the case under consideration 
the action will be subject to the relation
\be
{\bf L } = {\bf \ep} \bigg(
{}^{(n)}R - F_{\mu \nu}F^{\mu \nu} - \ga~ \ep^{a a_{1}a_{2} b_{1}b_{2} \dots m_{1}m_{2}}
A_{a}~F_{a_{1}a_{2}} \dots F_{m_{1}m_{2}} \bigg),
\label{lag}
\ee
where $\ga$ corresponds to CS coupling constant,
$ {\bf \ep}$ is the $n$-dimensional volume element,
$F_{\mu \nu} = 2\na_{[\mu}A_{\nu]}$. We remark that above Lagrangian applies to an odd dimensional spacetimes
$(n = 2d + 1)$ for which it includes the CS term.
\par
The equations of motion for the n-dimensional gravity with CS term yield
\ben \label{m1}
G_{\mu \nu} - T_{\mu \nu}(F) &=& 0, \\
\na_{\mu} F^{\mu \nu} &-&
 {\ga (n - 1) \over 4 } \ep^{\nu a_{1}a_{2} b_{1}b_{2} \dots m_{1}m_{2}}
~F_{a_{1}a_{2}} \dots F_{m_{1}m_{2}} = 0, \label{m2}         
\een
while the energy momentum tensor consist only of the Maxwell field contribution, i.e.,
\be
T_{\mu \nu}(F) =
2 F_{\mu \beta}F_{\nu}^{\beta} - {1 \over 2} g_{\mu \nu} F^2.
\ee
To deal with the problem of
the {\it physical version} of the first law of black rings thermodynamics 
we shall first begin with the explicit expressions
for the variation of mass and angular momentum and 
perform variation of the Lagrangian (\ref{lag}) evaluating
the variations of the adequate fields, which implies
\ben \label{dl}
\delta {\bf L} &=& {\bf \epsilon} \bigg(
G_{\mu \nu} - T_{\mu \nu}(F) \bigg)~ \delta g^{\mu \nu}
- \bigg( 4 \na_{\alpha} F^{\alpha \beta} - {\ga (n - 1) \over 4 }
 \ep^{\beta a_{1}a_{2} b_{1}b_{2} \dots m_{1}m_{2}}F_{a_{1}a_{2}} \dots F_{m_{1}m_{2}}
\bigg)~\delta A_{\beta} 
+ d {\bf \Theta}.
\een
In our paper we
denote fields in the underlying theory by $\psi_{\alpha}$,
while their variations by $\delta \psi_{\alpha}$. 
Having in mind 
relation (\ref{dl}) we get
the symplectic $(n - 1)$-form
$\Theta_{j_{1} \dots j_{n-1}}[\psi_{\alpha}, \delta \psi_{\alpha}]$ of the form as
\be
\Theta_{j_{1} \dots j_{n-1}}[\psi_{\alpha}, \delta \psi_{\alpha}] =
\ep_{\mu j_{1} \dots j_{n-1}} \bigg[
\omega^{\mu} - 
4 F^{\mu \beta} ~\delta A_{\beta}
- 2 \ga (n - 2) 
 \ep^{a \mu a_{2} \dots m_{1}m_{2}}~A_{a}~F_{a_{1}a_{2}} \dots F_{m_{1}m_{2}}
~\delta A_{a_{2}} 
 \bigg],
\ee 
where $\omega_{\mu} = \na^{\alpha} \delta g_{\alpha \mu} - \na_{\mu} 
\delta g_{\beta}{}{}^{\beta}.$
\par 
In the next step one ought to find
the Noether $(n - 1)$-form with respect to this above mentioned Killing vector.
Namely, we look for the form subject to the relation 
${\cal J}_{j_{1} \dots j_{n-1}} = \ep_{m j_{1} \dots j_{n-1}} {\cal J}^{m}
\big[\psi_{\alpha}, {\cal L}_{\xi} \psi_{\alpha}\big]$. Thus we have finally left with
\ben
{\cal J}_{j_{1} \dots j_{n-1}} &=& 
d \bigg( Q^{GR} + Q^{FCS} \bigg)_{j_{1} \dots j_{n-1}}
+ 2~ \ep_{\delta j_{1} \dots j_{n-1}} \bigg( G^{\delta}{}{}_{\eta} - T^{\delta}{}{}_{\eta}(F)
\bigg) \xi^{\eta} \\ \nonumber
&+& \ep_{m j_{1} \dots j_{n-1}}~ \xi^{d} A_{d}~ \bigg[
- 4 \na_{\beta} F^{\beta m} +
\ga (n - 1)
 \ep^{m a_{1} a_{2} b_{1} b_{2} \dots m_{1}m_{2}}~F_{a_{1}a_{2}}~F_{b_{1}b_{2}} \dots F_{m_{1}m_{2}} \bigg],
\een
where we denoted by $Q_{j_{1} \dots j_{n-2}}^{GR}$ the expression as follows:
\be
Q_{j_{1} \dots j_{n-2}}^{GR} = - \ep_{j_{1} \dots j_{n-2} a b} \na^{a} \xi^{b},
\ee
and by $Q_{j_{1} \dots j_{n-2}}^{FCS}$ the relation of the following form:
\be
Q_{j_{1} \dots j_{n-2}}^{FCS} = \ep_{m \delta j_{1} \dots j_{n-2}}
\bigg( {2 F^{\delta m} \over (n - 2)!} -  {\ga (n - 2) \over (n - 2)!}
\ep^{\delta m a b_{1} b_{2} \dots m_{1}m_{2}}~A_{a}~F_{b_{1}b_{2}} \dots F_{m_{1}m_{2}}
\bigg)~\xi_{d} A^{d}.
\ee
Having in mind that ${\cal J}[\xi] = dQ[\xi] + \xi^{\alpha} {\bf C}_{\alpha}$,
where ${\bf C}_{\alpha}$ is an $(n-1)$-form constructed from dynamical fields, i.e., from
$g_{\mu \nu}$ and $F_{\mu \nu}$ gauge field. One can identify
$Q_{j_{1} \dots j_{n-1}} = (Q^{GR} + Q^{F} + Q^{CS})_{j_{1} \dots j_{n-1}}$ with the 
Noether charge for the considered theory.
It reveals then that ${\bf C}_{\alpha}$ reduces to the following:
\ben
C_{a j_{1} \dots j_{n-1}} &=& 2 \ep_{m j_{1} \dots j_{n-1}}
\bigg[ G_{a}{}{}^{m} - T_{a}{}{}^{m}(F) \bigg] \\ \nonumber
&+&
\ep_{m j_{1} \dots j_{n-1}}~
A_{a} \bigg(
- 4 \na_{\beta} F^{\beta m} + \ga (n - 1) 
 \ep^{m a_{1} a_{2} b_{1} b_{2} \dots m_{1}m_{2}}~F_{a_{1} a_{2}}~F_{b_{1}b_{2}} \dots F_{m_{1}m_{2}}
\bigg).
\een
The case when ${\bf C}_{\alpha} = 0$ is responsible for the source-free Eqs. of motion but on the contrary,
when this is not the case, one gets the following:
\ben
G_{\mu \nu} - T_{\mu \nu}(F) &=& T_{\mu \nu}(matter) , \\
\na_{\beta} F^{\beta \mu} &=& {\ga (n - 1) \over 4}
 \ep^{\mu a_{1} a_{2} b_{1} b_{2} \dots m_{1}m_{2}}~F_{a_{1} a_{2}}~F_{b_{1}b_{2}} \dots F_{m_{1}m_{2}}
+ j^{\mu}(matter).
\een
Let us assume further that $(g_{\mu \nu}, F_{\alpha \beta})$ are solutions 
of source-free equations of motion and 
$(\delta g_{\mu \nu},~\delta F^{\alpha \beta})$
are the linearized perturbations satisfying Eqs. of motion with sources
$\delta T_{\mu \nu}(matter)$ and $ \delta j^{\mu}(matter)$. It
enables us to conclude that
\be
\delta  C_{a j_{1} \dots j_{n-1}} = \ep_{m j_{1} \dots j_{n-1}} \bigg(
2 \delta T_{a}{}{}^{m}(matter) + j^{m}(matter)A_{a} \bigg).
\ee
The Killing vector field $\xi_{\alpha}$ describes a symmetry of the background
matter field. By virtue of it the formula for a conserved quantity related with 
the Killing vector field $\xi_{\alpha}$ may be expressed as
\ben \label{hh}
\delta H_{\xi} &=& - \int_{\Sigma}\ep_{m j_{1} \dots j_{n-1}} \bigg[
2 \delta T_{a}{}{}^{m}(matter) \xi^{a} + 
 A_{a} \xi^{a}~\delta j ^{m }(matter) \bigg]
 \\ \nonumber
&+& \int_{\p \Sigma}\bigg[
\delta Q(\xi) - \xi \cdot \Theta \bigg].
\een
As in Ref.\cite{gao01}, if one takes $\xi^{\alpha}$ to be an asymptotic time translation $t^{\alpha}$, then
it enables us to identify
$M = H_{t}$ and in particular we obtain 
the variation of the ADM mass. Thus in this picture we get the following
\ben \label{mm}
\alpha~ \delta M &=& - \int_{\Sigma} \ep_{m j_{1} \dots j_{n-1}} \bigg[
2 \delta T_{a}{}{}^{m}(matter) t^{a} + 
A_{a} t^{a}~\delta j ^{m }(matter) \bigg] \\ \nonumber
&+& \int_{\p \Sigma}\bigg[
\delta Q(t) - t \cdot \Theta \bigg],
\een
where we denoted by $\alpha = {n-3 \over n-2}$.
On the other hand, if we turn our attention to
the Killing vector fields $\phi_{(i)}$ which are responsible
for the rotation in the adequate directions, one gets the relations for angular 
momenta written in the form as
\ben \label{jj}
\delta J_{(i)} &=& \int_{\Sigma} \ep_{m j_{1} \dots j_{n-1}} \bigg[
2 \delta T_{a}{}{}^{m}(matter) \phi_{(i)}^{a} + 
A_{a} \phi_{(i)}^{a}~\delta j ^{m }(matter) \bigg] \\ \nonumber
&+& \int_{\p \Sigma}\bigg[
\delta Q(t) - \phi_{(i)} \cdot \Theta \bigg].
\een
Next, we proceed to the {\it physical process} version of the first law of black ring thermodynamics.
Let us us assume that
$(g_{\mu \nu},~F_{\alpha \beta})$ are solutions to the source free
Einstein equations. Moreover, let $\eta_{\alpha}$ denotes
the event horizon Killing vector field of the form
\be
\eta^{\mu} = t^{\mu} + \sum_{i} \Omega_{(i)} \phi^{\mu (i)}.
\label{kil}
\ee
Suppose, now that one perturbs the black ring by dropping into it some matter. Furthermore,
assume that the black ring will be not destroyed in the process of it and it settles down to a stationary
solution \cite{gao01}. We shall compute changes of a mass and angular momenta using relations
(\ref{mm})-(\ref{jj}) and find the change of the horizon's area using
the Raychaudhuri equation.
In what follows we shall assume that
$\Sigma_{0}$ is an asymptotically flat
hypersurface which terminating on the black ring event horizon. Then, one takes into account 
the initial data on $\Sigma_{0}$ for a linearized perturbations of
$(\delta g_{\mu \nu},~ \delta F_{\alpha \beta})$
with $\delta T_{\mu \nu}(matter)$ and $\delta j^{\mu}(matter)$. We 
require that $\delta T_{\mu \nu}(matter)$ and $\delta j^{\mu}(matter)$
vanish
at infinity and the initial data for 
$(\delta g_{\mu \nu},~ \delta F_{\alpha \beta})$
disappear in the vicinity of the black ring horizon $\cal H$ on 
the hypersurface $\Sigma_{0}$. 
These requirements provide
that for the initial time
$\Sigma_{0}$, the considered black hole is unperturbed. 
Hence the perturbations vanish near the internal boundary $\p \Sigma_{0}$,
one gets from relations (\ref{mm}) and (\ref{jj}) that the following is fulfilled:
\ben \label{ppp}
\alpha~ \delta M &-&  \sum_{i} \Omega_{(i)} \delta J^{(i)} = \\ \nonumber
&-& \int_{\Sigma_{0}} \ep_{m j_{1} \dots j_{n-1}} \bigg[
2 \delta T_{a}{}{}^{m}(matter) \eta^{a} +  \eta^{a} A_{a}
~\delta j ^{m}(matter) \bigg] \\ \nonumber
&=& \int_{\cH} \gamma ^{\alpha}~k_{\alpha}~\bep_{j_{1} \dots j_{n-1}},
\een
where $\bep_{j_{1} \dots j_{n-1}} = n^{\delta}~\ep_{\delta j_{1} \dots j_{n-1}}$ while
$n^{\delta}$ is a future directed unit normal to the hypersurface $\Sigma_{0}$.
On the other hand,
$k_{\alpha}$ is tangent vector to the affinely parametrized null geodesics generators of the 
black ring event horizon.
\par
In the last term of relation (\ref{ppp}) we have replaced $n^{\delta}$ for
$k_{\delta}$. It can be done because of the fact that the current $\gamma^{\alpha}$ 
is conserved as well as by the assumption that all
the matter falls into black ring. \\
Before considering the integrals over the
black ring event horizon let take into account the following relation:
\ben \label{cf}
\cL_{\eta} A_{c}~\ep^{m c b_{1} b_{2} \dots m_{1}m_{2}}~F_{b_{1}b_{2}} \dots F_{m_{1}m_{2}}
&-& \eta^{k}F_{kc}~\ep^{m c b_{1} b_{2} \dots m_{1}m_{2}}~F_{b_{1}b_{2}} \dots F_{m_{1}m_{2}} 
\\ \nonumber
&=& \na_{c} \bigg(
A_{j }\eta^{j}~\ep^{m c b_{1} b_{2} \dots m_{1}m_{2}}~F_{b_{1}b_{2}} \dots F_{m_{1}m_{2}} \bigg).
\een
Because of the fact that $\eta_{\alpha}$ is
symmetry of the background solution
the first term of the left-hand side of Eq.(\ref{cf}) is equal to zero.
Then, let us turn to $n$-dimensional Raychaudhuri 
equation of the form as follows:
\be
{d \theta \over d \lambda} = - {\theta^{2} \over (n - 2)} - \sigma_{ij} \sigma^{ij}
- R_{\mu \nu} \xi^{\mu} \xi^{\nu},
\label{ray}
\ee
where $\lambda$ is an affine parameter corresponding to vector $k_{\alpha}$, $\theta$ is the expansion and
$\sigma_{ij}$ is shear. 
Shear and expansion 
vanish in the stationary background. 
Inspection of Eq.(\ref{ray}) provides that
$R_{\alpha \beta} k^{\alpha} k^{\beta} \mid_{\cH} = 0$, which in turn implies 
that $F_{\mu \alpha} F_{\nu}{}{}^{\alpha}k^{\mu}k^{\nu} \mid_{\cH} = 0$.
By the antisymmetry of the $U(1)$-gauge field tensor one reveals that 
$F_{\alpha \beta}k^{\alpha} \sim k_{\beta}$. 
It provides the pull-back of $F_{\alpha \beta}k^{\alpha}$ to the black hole ring
horizon vanishes. 
On the other hand this fact leads immediately to the conclusion that $F_{\alpha \beta}k^{\alpha}$ is a closed
one-form. Using the Hodge decomposition theorem it may be written as a sum
of closed and harmonic form. Due to the fact that the field Eqs. are satisfied the exact 
form is equal to zero and the only contribution stems from the harmonic part of the 
considered one-form. As in Refs.\cite{cop05,rog05br,rog05br1} 
by means of the duality between homology and cohomology
it follows that the surface terms will be of the form of a constant relating to the harmonic 
part of the one-form and the variation of a local charge.
We arrive at he following:
\be
\alpha~ \delta M -  \sum_{i} \Omega_{(i)} \delta J^{(i)} 
+ \Phi_{e}~\delta Q_{e} + \Phi_{m}~\delta q_{m} =
2 \int_{\cH}
\delta T_{\mu}{}{}^{\nu}(matter) \xi^{\mu} k_{\nu}.
\label{rh}
\ee
The right-hand side of Eq.(\ref{rh}),
can be found by the same procedure as described in Refs.\cite{gao01,rog02,rog05}, i.e.,
having in mind $n$-dimensional Raychauduri Eq. and 
using the fact that
the null generators of the event horizon of the perturbed black ring coincide with
the null generators of the unperturbed stationary black ring, one arrives at the expression
\be
\kappa~ \delta A = \int_{\cal H} \delta
T^{\mu}{}{}_{\nu}(matter) \xi^{\nu} k_{\mu},
\ee
where $\kappa$ is the surface gravity.\\
The {\it physical process} version of the first law of black rings mechanics may be written as
\be
\alpha~ \delta M -  \sum_{i} \Omega_{(i)} \delta J^{(i)} 
+ \Phi_{e}~\delta Q_{e} + \Phi_{m}~\delta q_{m} = 2 \kappa~ \delta A.
\ee

\section{Equilibrium state version of the first law of black ring mechanics}
In this section we shall derive the first law of black rings dynamics in n-dimensional
CS gravity by choosing an arbitrary cross section of the event horizon to the future
of the bifurcation sphere. As was remarked in \cite{gao03} this attitude enables one to 
treat fields which are not necessarily smooth through the horizon. The only requirement is that the pull-back
of these fields in the future of the bifurcation surface be smooth. Let us consider
the spacetime with asymptotic conditions at infinity and equipped with the Killing vector field
$\xi_{\mu}$, which introduces an asymptotic symmetry. It is turned out that there exists a conserved quantity
$H_{\xi}$ \cite{wal00}, which yields
\be
\delta H_{\xi} = \int_{\infty} \bigg( \bdel Q(\xi) - \xi \Theta \bigg).
\label{qua}
\ee
$\bdel$ is the variation which has no effect on $\xi_{\alpha}$ because of the fact that the Killing
vector field is treated as a fixed background and it ought not to be varied in expression (\ref{qua}).\\ 
In our considerations we were bound to the case of stationary axisymmetric black ring solution so the 
Killing vector field will be given by Eq.(\ref{kil}). We shall consider an asymptotically hypersurface 
$\Sigma$ ending on the part of the event horizon $\cal H$ to the future of the bifurcation surface. 
The cross section of the black ring horizon will constitute the inner boundary of the hypersurface
$\Sigma$. It will be denoted by $S_{\cal H}$. In our considerations of the first law of black ring 
dynamics we shall compare variations between two neighbouring states of the considered object.
In general, there is a freedom which points can be chosen to correspond when one compares two
slightly different solutions. In what follows we choose this freedom \cite{bar73}
to make $S_{\cal H}$ the same of the two solutions (freedom of the general coordinate transformation)
as well as we consider the case when the null vector remains normal to $S_{\cal H}$. Of course,
the stationarity and axisymmetricity of the solution will be preserved which in turn
causes that $\delta t^{\alpha}$ and $\delta \phi^{\mu (i)}$ will be equal to zero.
It yields that the variation of the Killing vector field $\eta_{\alpha}$ is of the form
$\delta \eta^{\mu} = \sum_{i} \delta \Omega_{(i)} \phi^{\mu (i)}$.
\par
As in the previous section let us assume that 
$(g_{\mu \nu}, F_{\alpha \beta})$ are solutions 
of the equations of motion and 
$(\delta g_{\mu \nu},~\delta F^{\alpha \beta})$
are the linearized perturbations satisfying Eqs. of motion. We shall
require that the pull-back of $F_{\alpha \beta}$ to the future of the bifurcation surface be smooth,
but not necessary smooth on it. We also assume that the fields and their variations fall off sufficiently
rapid at infinity. Having it all in mind we can write the relation
\be
\alpha~ \delta M -  \sum_{i} \Omega_{(i)} \delta J^{(i)} = 
\int_{S_{\cal H}} \bigg( \bdel Q(\eta) - \eta \Theta \bigg).
\ee
The same arguments as quoted in the previous section help us to conclude that the
following is satisfied:
\ben
\int_{S_{\cal H}} Q_{j_{1} \dots j_{n-2}}^{FCS}(\eta) &=& 
\int_{S_{\cal H}} Q_{j_{1} \dots j_{n-2}}^{F}(\eta) + \int_{S_{\cal H}} Q_{j_{1} \dots j_{n-2}}^{CS}(\eta) \\ \nonumber
&=&
\Phi_{e}~Q_{e} + \Phi_{m}~q_{m}.
\een
The variation $\bdel$ of $Q_{j_{1} \dots j_{n-2}}^{F}(\eta)$ implies
\ben \label{bar}
\bdel \int_{S_{\cal H}} Q_{j_{1} \dots j_{n-2}}^{F} (\eta) &=&
\delta \int_{S_{\cal H}} Q_{j_{1} \dots j_{n-2}}^{F} (\eta) -
\int_{S_{\cal H}} Q_{j_{1} \dots j_{n-2}}^{F} (\delta \eta) = 
\delta \bigg( \Phi_{e}~Q_{e} \bigg) \\ \nonumber 
&+&
{2 \over (n - 2)!}\int_{S_{\cal H}} \sum_{i} \delta \Omega_{(i)} \phi^{\mu (i)}
A_{\mu}~ 
\ep_{m \alpha j_{1} \dots j_{n-2}}~F^{m \alpha}.
\een
We can express the volume element $\ep_{\mu a j_{1} \dots j_{n-2}}$ by the volume element on $S_{\cal H}$ and
by the vector $N^{\alpha}$, which is {\it ingoing} future directed null normal to $S_{\cal H}$, together
with the  normalization
$N^{\alpha} \eta_{\alpha} = - 1$. 
It can be easily verify that this relation and Eq.(\ref{bar}) enables us to write
\ben \label{pt}
 \delta \Phi_{e}~Q_{e} &=& {4 \over (n - 2)!}
\int_{S_{\cal H}}\ep_{j_{1} \dots j_{n-2}}~F^{m \alpha}
~N_{m}~\eta_{\alpha}~\eta^{d} \delta A_{d}~ \\ \nonumber
&-& {2 \over (n - 2)!}\int_{S_{\cal H}}
\ep_{m \alpha j_{1} \dots j_{n-2}}~F^{m \alpha}
~ \sum_{i} \delta \Omega_{(i)} \phi^{\mu (i)}
A_{\mu}.
\een
We take now into account symplectic $(n-1)$-form for the potential
$A_{\nu}$. Due to the fact
that on the event horizon of black ring 
$F_{\mu \alpha} \eta^{\mu} \sim \eta_{\alpha}$
and expressing the volume element $\ep_{\mu a j_{1} \dots j_{n-2}}$ in the same form
as in the above case, one gets
\be
\int_{S_{\cal H}} \eta^{j_{1}}~\Theta_{j_{1} \dots j_{n-1}}^{F} =
{4 \over (n - 2)!}
\int_{S_{\cal H}} \ep_{j_{1} \dots j_{n-2}}~ F^{\delta \alpha }~N_{\delta}~
\eta_{\alpha}~\delta A_{d} \eta^{d}.  
\label{bb1}
\ee
Using Eq.(\ref{pt}) together with the expressions (\ref{bb1}) 
we finally conclude
\be
\bdel \int_{S_{\cal H}} Q_{j_{1} \dots j_{n-2}}^{F} (\eta)
- \eta^{j_{1}}~\Theta_{j_{1} \dots j_{n-1}}^{F} = 
\Phi_{e}~\delta Q_{e}.
\label{char}
\ee
The same procedure as above applied to Chern-Simons term provides the following:
\ben \label{ss}
\bdel \int_{S_{\cal H}}~ Q_{j_{1} \dots j_{n-2}}^{CS} (\eta) 
&=& \delta \bigg( \Phi_{m}~q_{m} \bigg) \\ \nonumber
&-& \int_{S_{\cal H}} \ep_{m \alpha j_{1} \dots j_{n-2}} 
{\ga (n - 2) \over (n - 2)!}
 \ep^{m \alpha a b_{1} b_{2} \dots m_{1}m_{2}}~A_{a}~F_{b_{1}b_{2}} \dots F_{m_{1}m_{2}}~
\sum_{i} \delta \Omega_{(i)} \phi^{d (i)}~A_{d},
\een
while variation of the potential multiplied by the local charge gives us
\ben
\delta \Phi_{m} ~q_{m} &=& \int_{S_{\cal H}} \ep_{j_{1} \dots j_{n-2}} 
{2 \ga (n - 2) \over (n - 2)!}~
\ep^{m \alpha a b_{1} b_{2} \dots m_{1}m_{2}}~N_{\alpha}~\eta_{m}~A_{a}~F_{b_{1}b_{2}} \dots F_{m_{1}m_{2}}~
\eta^{d} \delta A_{d} \\ \nonumber
&+&
\int_{S_{\cal H}}
\ep_{m \alpha j_{1} \dots j_{n-2}} 
{\ga (n - 2) \over (n - 2)!}
\ep^{m \alpha a b_{1} b_{2} \dots m_{1}m_{2}}~A_{a}~F_{b_{1}b_{2}} \dots F_{m_{1}m_{2}}
\sum_{i} \delta \Omega_{(i)} \phi^{d (i)}~A_{d}.
\een
Consequently it is easy to see that
\be
\int_{S_{\cal H}} \eta^{j_{1}}~\Theta_{j_{1} \dots j_{n-1}}^{CS} =
\int_{S_{\cal H}} \ep_{j_{1} \dots j_{n-2}} 
{2 \ga (n - 2) \over (n - 2)!}~
 \eta_{m}~\ep^{m j a b_{1} b_{2} \dots m_{1}m_{2}}~N_{j}~A_{a}~F_{b_{1}b_{2}} \dots F_{m_{1}m_{2}}~
\eta^{d}~ \delta A_{d}.
\label{ss1}
\ee
By virtue of Eq.(\ref{ss}) and (\ref{ss1}) we can finally conclude the following:
\be
\bdel \int_{S_{\cal H}} Q_{j_{1} \dots j_{n-2}}^{CS} (\eta)
- \xi^{j_{1}}~\Theta_{j_{1} \dots j_{n-1}}^{CS} = 
\Phi_{m}~\delta q_{e}.
\label{char2}
\ee
Consider, next, the contribution connected with gravitational
field \cite{bar73}. It implies
\be
\int_{S_{\cal H}} Q_{j_{1} \dots j_{n-2}}^{GR} (\eta) = 2 \kappa A,
\ee
where $A = \int_{S_{\cal H}} \ep_{j_{1} \dots j_{n-2}}$ 
is the area of the black ring horizon. In terms of the above derivations one obtains 
\be
\bdel \int_{S_{\cal H}} Q_{j_{1} \dots j_{n-2}}^{GR} (\eta) =
2 \delta \bigg( \kappa A \bigg) + 2 \sum_{i} \delta \Omega_{(i)}~ J^{(i)},
\label{gr1}
\ee
where $J^{(i)}= {1 \over 2}\int_{S_{\cal H}}\ep_{j_{1} \dots j_{n-2} a b} \na^{a} \phi^{(i)b}$ 
is the angular momentum connected with the Killing vector
field $\phi_{(i)}$ responsible for the rotation in the adequate directions.
Conducting the calculations as in Ref.\cite{bar73}
it could be found that the integral over the black ring horizon from the product
$\xi^{j_{1}}~ \Theta_{j_{1} \dots j_{n-1}}^{GR} (\eta)$ can be written as
\be
\int_{S_{\cal H}} \eta^{j_{1}}~ \Theta_{j_{1} \dots j_{n-1}}^{GR} (\eta) =
2 A~ \delta \kappa + 2 \sum_{i} \delta \Omega_{(i)}~ J^{(i)}.
\label{gr2}
\ee
From Eq.(\ref{gr1}) and (\ref{gr2}) one imediatelly obtains the expression
\be
\bdel \int_{S_{\cal H}} Q_{j_{1} \dots j_{n-2}}^{GR} (\eta)
- \xi^{j_{1}}~\Theta_{j_{1} \dots j_{n-1}}^{GR} = 
2 \kappa~\delta A.
\label{arr}
\ee
Summing it all up, namely taking into account Eqs.(\ref{char}) and (\ref{arr}),
we find that 
the {\it equilibrium state} version of the first law of black rings mechanics in Einstein
$n$-dimensional gravity with CS term, can be determined by the formula:
\be
\alpha~ \delta M -  \sum_{i} \Omega_{(i)} \delta J^{(i)} 
+ \Phi_{e}~\delta Q_{e} + \Phi_{m}~\delta q_{m} 
= 2 \kappa~\delta A.
\ee

\section{Conclusions}
This paper provides the first law of black ring thermodynamics both for the {\it physical process}
version and {\it equilibrium state} one. We considered $n$-dimensional gravity with CS term
being the simplest generalization of the minimal five-dimensional supergarvity
in which a stationary black rings solutions were achieved. In five-dimensional case the solutions were 
sufficiently complicated that a first law could not be obtained by inspection (in Ref.\cite{cop05} 
the first law of black ring thermodynamics was found by means of the ADM formalism method \cite{sud92}).
\par
Deriving the {\it physical process} version we changed infinitesimally a stationary black ring 
solution by throwing matter into it. Taking into account that the black ring settles down to a stationary
state we derive the first law of thermodynamics. The same form of the law we obtained
considering {\it equilibrium state} version. We choose an arbitrary cross section
of the black ring event horizon to the future of bifurcation surface, contrary to the previous 
derivations based on taking into considerations bifurcation surface as the boundary of the
hypersurface extending to spatial infinity.
In general, as was remarked in Ref.\cite{gao03} this attitude enables one to 
treat fields which are not necessary smooth through the event horizon. The only requirement 
is that their pull-back in the future of the bifurcation surface be smooth.





\end{document}